\documentstyle[aps,prl]{revtex}

\begin{document}

\input epsf.sty
\twocolumn[\hsize\textwidth\columnwidth\hsize\csname %
@twocolumnfalse\endcsname

\draft

\widetext

\title{NEUTRON SCATTERING STUDY OF ELASTIC MAGNETIC SIGNALS
IN SUPERCONDUCTING La$_{1.94}$Sr$_{0.06}$CuO$_{4}$}

\author{S. Wakimoto,$^{\dag}$ K. Yamada,$^{\dag}$\footnote{\tiny
Present address: Institute for Chemical Research, Kyoto University, Gokasho, Uji
610-0011, Japan} S. Ueki,$^{\dag}$ G. Shirane,$^{\ddag}$
Y. S. Lee,$^{\S}$ S. H. Lee,$^{\P,\parallel}$ M. A. Kastner,$^{\S}$ K. Hirota,$^{\dag}$
P. M. Gehring,$^{\P}$ Y. Endoh,$^{\dag}$ and R. J. Birgeneau$^{\S}$}
\address{$^{\dag}$Department of Physics, Tohoku University, Sendai 980-8578, Japan}
\address{$^{\ddag}$Department of Physics, Brookhaven National Laboratory, Upton, NY
11973, USA}
\address{$^{\S}$Department of Physics, Massachusetts Institute of Technology,
Cambridge, MA 02139, USA}
\address{$^{\P}$NIST Center for
Neutron Research, Gaithersburg MD 20899, USA}
\address{$^{\parallel}$University of Maryland, College Park, MD 20742, USA}

\date{\today}
\maketitle

\vspace{-0.1in}

\begin{abstract}

Recent neutron-scattering experiments on La$_{2-x}$Sr$_{x}$CuO$_{4}$
single crystals by Wakimoto {\it et al.} have revealed that elastic magnetic
peaks appear at low temperatures in both insulating $(x=0.02-0.05)$ and
superconducting $(x=0.06)$ samples.
We have carried out further investigations particularly on the elastic
incommensurate peaks for $x=0.06$, and found that the integrated intensity
drastically changes across the low temperature insulator-superconductor boundary;
the intensity of $x=0.06$ is 4 times smaller than that of $x=0.05$, while the
intensity in the insulating region stays constant.

\end{abstract}

\vspace{0.5cm}
{\it Keywords:} A. Superconductors, B. Crystal growth, C. Neutron scattering,
D. Magnetic properties, Superconductivity 

\pacs{PACS numbers: 74.72.Dn, 75.10.Jm, 75.50.Ee, 71.45.Ln, 75.70.Kw}

\phantom{.}
]
\narrowtext

%
%
%

The 2-1-4 type cuprate shows various magnetic and electronic properties
with changing charge carriers introduced in the CuO$_{2}$ planes.
In  the La$_{2-x}$Sr$_{x}$CuO$_{4}$ system, long-range three-dimensional (3D)
antiferromagnetic (AF) order destroyed at the hole concentration $x=0.02$
and spin-glass (SG) behavior appears in the region of $0.02 \leq x \leq 0.05$.
With further doping, the system shows
superconductivity for $0.06 \leq x \leq 0.25$  at low temperatures.
Throughout the superconducting region, inelastic magnetic peaks are observed
at incommensurate positions around $(\pi, \pi)$ by neutron scattering experiments\cite{K.Yamada_98},
suggesting  strong correlations between the dynamical spin fluctuations and the superconductivity.
On the other hand, elastic magnetic peaks are observed at low temperatures
in the both the insulating and superconducting regions.
In the superconducting region, sharp-$q$ elastic peaks are observed at incommensurate positions
for $x=0.12$,
where the superconductivity is slightly suppressed (1/8 problem).\cite{T.Suzuki_98,H.Kimura_98}
In the insulating region, Keimer {\it et al.}\cite{B.Keimer_92} and Wakimoto {\it et al.}
\cite{waki_neut} observed a commensurate magnetic peak.
Wakimoto {\it et al.}\cite{waki_neut} established that the elastic peak exhibits
fundamental changes
at the insulator-superconductor boundary; the elastic commensurate
peak becomes broader with increasing $x$ in the insulating region and rapidly sharpens 
in the superconducting region.
They also discovered new incommensurate elastic peaks at $x=0.05$, 
near the lower critical concentration for superconductivity, which positions are
45$^{\circ}$ rotated around $(\pi, \pi)$ from those of superconducting samples.
Although such remarkable changes appear at the boundary, the SG transition temperatures
determined by systematic $\mu$SR study\cite{Ch.Niedermayer_98}
appear to fall on a continuous universal curve for $0.02 \leq x \leq 0.10$.

It becomes increasingly important to study the elastic component systematically
from the insulating to the superconducting region in order to clarify
the relation between the static magnetic correlations and the superconductivity.
We have performed systematic neutron-scattering experiments
for $0.03 \leq x \leq 0.06$, in particular focusing on changes
of the magnetic properties at the insulator-superconductor boundary.
In the present paper, we report results for $x=0.06$ and discuss the $x$-dependence
of the elastic peak intensity referring to the results for $x=0.03, 0.04$ and $0.05$
reported in Ref.\cite{waki_neut}.

A single crystal of $x=0.06$, whose size is 6~mm in diameter and 25~mm in length, was grown by
an improved travelling-solvent floating-zone method\cite{lee_98} at Tohoku University.
The growth conditions are the same as those used for the 
$x=0.03, 0.04$ and $0.05$ crystals reported in Ref.\cite{waki_neut}.
We annealed the $x=0.06$ crystal in flowing oxygen at 900 $^{\circ}$C for
12 hours before experiments.
We measured the Meisner signal using a SQUID magnetometer
with the applied field of 10 Oe parallel to the CuO$_{2}$ plane after a cooling process in zero field.
Neutron-scattering experiments were performed on the HER cold neutron
triple-axis spectrometer located at the JAERI JRR-3M reactor.
To study the elastic component in the $x=0.06$ crystal,
we utilized an incident neutron energy of 5~meV to obtain high energy resolution
($\sim 0.25$~meV) together with a Be filter to exclude higher order contaminations.
The crystal was oriented so as to give the $(h\ k\ 0)$ zone,
and sealed in an Al can filled with He gas. In the present paper, we use the tetragonal
$I4/mmm$ crystallographic notation.

\begin{figure}
\centerline{\epsfxsize=2.4in\epsfbox{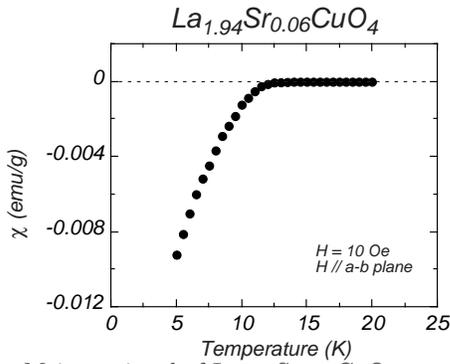}}
\caption{Meisner signal of La$_{1.94}$Sr$_{0.06}$CuO$_{4}$ measured by a SQUID 
magnetometer. The measurement was performed in a magnetic field of 10 Oe  
parallel to the CuO$_{2}$ planes after zero-field cooling.}
\label{Fig:meisner}
\end{figure}

Figure \ref{Fig:meisner} shows the Meisner signal of the $x=0.06$ crystal. 
The onset temperature of superconductivity \linebreak

\noindent
$T_{c}$(onset)
is 12~K which is consistent with the $x$-dependence of $T_{c}$(onset) reported in
Ref.\cite{K.Yamada_98}, suggesting that the Sr concentration of the crystal is indeed $0.06$.

\begin{figure}
\centerline{\epsfxsize=2.4in\epsfbox{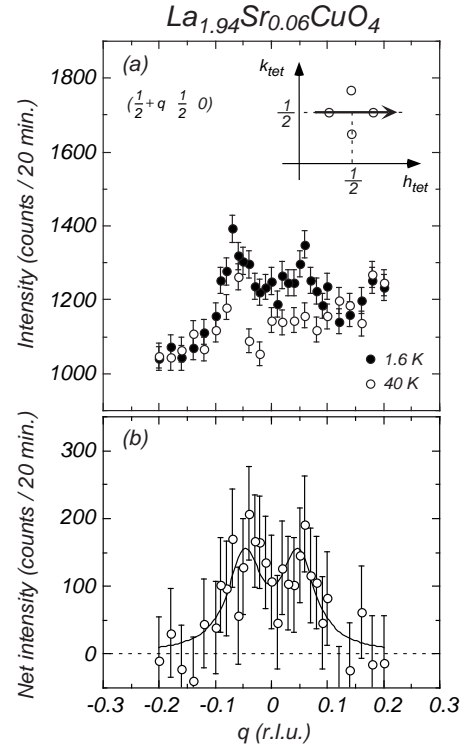}}
\vspace{2mm}
\caption{(a) Peak profiles measured using the HER spectrometer along the scan
trajectory shown in the inset figure. Closed circles are data at 1.6 K and open
circles are data at 40 K.  (b) Net intensity calculated by subtracting the 40 K
data from the 1.6 K data. The solid line is the result of fitting using a double Lorentzian function.}
\label{Fig:profile}
\end{figure}

Figure \ref{Fig:profile} indicates peak profiles along the scan
trajectory shown in the inset. In Fig.\ref{Fig:profile}(a), closed and open symbols indicate
data at 1.6~K and 40~K respectively. Although the data at 40~K are contaminated around $q=-0.06$,
there exist clear differences between the profiles at the two temperatures.
To analyze the incommensurate elastic peaks excluding the contamination,
we calculated the net intensity by subtracting the 40~K data from the 1.6~K data.
Figure \ref{Fig:profile}(b) shows the resultant net intensity. The solid line is the result of fitting
using a double Lorentzian function consisting of two Lorentzians which are symmetric around $q=0$.
From the fitting result, the incommensurability of the elastic incommensurate peaks
is $0.048(\pm 0.005)$ r.l.u., which is very close to that of the inelastic incommensurate
peaks.\cite{K.Yamada_98}

\begin{figure}
\centerline{\epsfxsize=2.4in\epsfbox{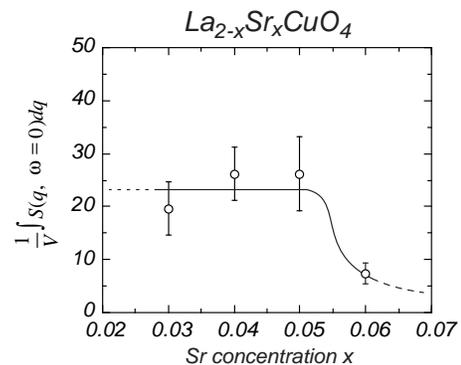}}
\vspace{2mm}
\caption{Sr concentration $x$ dependence of the normalized integrated intensity
of the magnetic elastic peaks which was calculated by summation of the total integrated peak
intensity around $(\pi, \pi)$ and normalized by the sample volume.
The solid line is a guide to the eye.}
\label{Fig:integrated}
\end{figure}

To investigate changes of the magnetic properties across the boundary between the insulating
and superconducting region, we plot the $x$-dependence of the integrated intensity of elastic
magnetic peaks as shown in Fig.\ref{Fig:integrated}. 
The horizontal axis indicates the integrated intensity normalized by sample volume.
The integration was carried out for all elastic magnetic peaks around $(\pi, \pi)$.
Note that the data for $x=0.03, 0.04$ and $0.05$ referred from Ref.\cite{waki_neut}
are calculated from the results
obtained using the SPINS spectrometer in the identical instrumental configuration.
Since the HER data of the $x=0.05$ crystal in the identical instrumental configuration used in
the present study are also reported in Ref.\cite{waki_neut},
we can make the necessary corrections for the $x=0.06$ data.
As clearly shown in Fig.\ref{Fig:integrated}, the elastic magnetic
intensity suddenly decreases across the phase boundary at $x_{c} \sim 0.06$. 
This fact strongly suggest a competing relation between the superconductivity
and the static magnetic correlations.\linebreak


\noindent
On the other hand, the integrated intensity of the elastic magnetic peaks in the $x=0.12$
sample
is half of that for $x=0.06$ while the peak shape is clearly different
from that of the $x=0.06$ sample; the elastic peaks in the $x=0.12$ sample are sharper than
those for $x=0.06$.\cite{H.Kimura_98}
Note that the onset $T_{c}$ at $x=0.12$ is 31~K, higher than that of $x=0.06$.
Taking account of this difference of $T_{c}$, we speculate that the magnetic moment which contributes
to the elastic magnetic peaks decreases as $T_{c}$ increases
and the static correlation length becomes larger near $x=1/8$.
However it is necessary to study systematically the behavior in the intermediate concentrations
between $x=0.06$ and $0.12$ to clarify the relation between the superconductivity
and the static magnetic correlations in more detail.

We gratefully acknowledge H. Fukuyama, J. M. Tranquada, V. Emery and H. Kimura for their
invaluable discussions and suggestions. The present work was supported by the US-Japan
Cooperative Research Program on Neutron Scattering.  The work at Tohoku was supported
by a Grant-in-Aid for Scientific Research of Monbusho and the Core Research for Evolutional
Science and Techonology (CREST) Project sponsored by the Japan Science and Technology Corporation.
The work at MIT was supported by the NSF under Grant No.\ DMR97-04532 and by the MRSEC Program
of the National Science Foundation under Award No.\ DMR94-00334.
The work at Brookhaven National Laboratory was carried out under Contract No.\ DE-AC02-98CH10886,
Division of Material Science, U. S.  Department of Energy.
The work at SPINS is based upon activities supported by the National Science
Foundation under Agreement No. DMR-9423101.


\begin{references}

\bibitem{K.Yamada_98} K. Yamada, C. H. Lee, K. Kurahashi, J. Wada, S. Wakimoto,
S. Ueki, H. Kimura, Y. Endoh, S. Hosoya, G. Shirane, R. J. Birgeneau, M. Greven,
M. A. Kastner, and Y. J. Kim, Phys.\ Rev.\ B {\bf 57}, 6165 (1998).

\bibitem{T.Suzuki_98} T. Suzuki, T. Goto, K. Chiba, T. Shinoda, T. Fukase, H.
Kimura, K. Yamada, M. Ohashi, and Y. Yamaguchi, Phys.\ Rev.\ B {\bf 57}, 3229
(1998).

\bibitem{H.Kimura_98} H. Kimura, K. Hirota, H. Matsushita, K. Yamada, Y. Endoh,
S. H. Lee, C. F. Majkrzak, R. Erwin, G. Shirane, M. Greven, Y. S. Lee, M. A.
Kastner, and R. J. Birgeneau, Phys.\ Rev.\ B (to be published).

\bibitem{B.Keimer_92} B. Keimer, N. Belk, R. J. Birgeneau, A. Cassanho, C. Y.
Chen, M. Greven, M. A. Kastner, A. Aharony, Y. Endoh, R. W. Erwin, and G.
Shirane, Phys.\ Rev.\ B {\bf 46}, 14034 (1992).

\bibitem{waki_neut} S. Wakimoto, R. J. Birgeneau, Y. Endoh, P. M. Ghering, K. Hirota,
M. A. Kastner, S. H. Lee, Y. S. Lee, G. Shirane, S. Ueki and K. Yamada, unpublished.

\bibitem{Ch.Niedermayer_98} Ch. Niedermayer, C. Bernhard, T. Blasius, A. Golnik,
A. Moodenbaugh, and J. I. Budnick, Phys.\ Rev.\ Lett. {\bf 80}, 3843 (1998).

\bibitem{lee_98} C. H. Lee, N. Kaneko, S. Hosoya, K. Kurahashi, S. Wakimoto, K. Yamada
and Y. Endoh, Supercond.\ Sci.\ Technol. {\bf 11}, 891 (1998).

\end{references}
\end{document}